\documentclass[preprint,authoryear,12pt]{elsarticle}
\usepackage[mathscr]{eucal}
\usepackage{graphics,pstricks,pst-node,clrscode,longtable,pstricks-add,epsfig}
\usepackage{clrscode}
\usepackage{pbsi}

\usepackage{amssymb}

\journal{Transportation Research Part C: Emerging Technologies}

\begin{document}

\begin{frontmatter}

\title{Partial order approach to compute shortest paths in  multimodal networks}
\author{Andrew Ensor\corref{cor2}}
\ead{andrew.ensor@aut.ac.nz}
\author{Felipe Lillo\corref{cor1}}
\ead{flillo@ucm.cl}

\cortext[cor2]{Tel:+64 9 921 9999 extn:8485}
\cortext[cor1]{Tel:+64 9 921 9999 extn:8953}

\address{School of Computing and Mathematical Sciences,
Auckland University of Technology,
2-14 Wakefield St, Auckland 1142, New Zealand.}

\begin{abstract}
Many networked systems involve multiple modes of transport. Such
systems are called multimodal, and examples include logistic
networks, biomedical phenomena, manufacturing process and
telecommunication networks.
Existing techniques for determining optimal paths in
multimodal networks have either required
heuristics or else application-specific constraints to obtain
tractable problems, removing the multimodal traits of the network during analysis.
In this paper weighted coloured--edge graphs are introduced to
model multimodal networks, where colours represent the modes of
transportation.
Optimal paths are selected using a partial order that compares the
weights in each colour, resulting in a Pareto optimal set of shortest paths.
This approach is shown to be tractable through experimental
analyses for random and real multimodal networks without the need
to apply heuristics or constraints.
\end{abstract}


\end{frontmatter}

\newcommand{\bin}[2]{
                      \left(
                            \begin{array}{@{}c@{}}
                                #1 \\ #2
                            \end{array}
                      \right)                   }


\section{Introduction}\label{Se:Introduction}
Extensive scientific literature has been devoted in the last three
decades to the study of multimodal networks (MMN).
During this time, research has mainly focused on practical
applications for freight or urban transportation
(see \cite{D937} and \cite{D949} for extensive reviews).
As a system in which several means of transport are available,
a multimodal system is able to emulate a wide spectrum of real life
phenomena beyond the field of logistics.
Areas such as computer networks, biology and
manufacturing have begun to utilize multimodal networks for
studying and modelling situations. Examples can be found in
papers by \cite{D92}, \cite{D920}, \cite{D933}, \cite{D941},
\cite{D957}, \cite{D950} and \cite{D964}.

From a modelling perspective, a range of techniques have been
used to model MMN. They can be loosely classified into three
predominant domains: mathematical programming, weighted graphs and
multi--weighted graphs.

\subsection{The mathematical programming approach}\label{SubSe:MathematicalProgrammingApproach}

These techniques are characterized by making use of linear or
non--linear formulations for representing a MMN by a set of
equations.

Linear programming techniques are suitable when each decision
variable is a linear combination of the problem parameters,
\cite{D935}. Integer programming and mixed integer programming
stand out as the most common linear programming techniques used
for multimodal modelling. Sample papers using linear programming
as a modelling tool are given by \cite{D952} and \cite{Kim1999}.

Non-linear programming is another renowned modelling technique for
MMN. It is mainly used to build intricate cost functions, and
principally deals with second order equations satisfying convex or
concave properties. Examples are provided by \cite{D942}, \cite{D936} and
\cite{D916} which have opted to use non-linear programming as
their main modelling approach.

In the  mathematical programming approach, mode options are
visualized as decision variable indices, which considerably
increases the complexity of the problem. Relaxation  or cutting
plane techniques are commonly used to make the problem tractable.
Interesting papers tackling general views of mathematical
programming for intermodal transportation (the transportation of
goods) and urban transportation were by \cite{D937} and \cite{D955}
respectively.

\subsection{The weighted graph approach}\label{SubSe:WeightedGraphApproach}

In this approach, a node typically represents a location, such as
a warehouse, transportation hub or network router, and an edge
represents a transportation link, such as rail line, a bus or a
wireless connection. A variety of graphs have been used to study
these transport systems, such as digraphs, multigraphs,
hypergraphs and grid graphs. \cite{Ayed2008} provides a
general classification for MMN models based on weighted graph
approaches. In particular, the article emphasizes the use of
multigraphs, in which there might be multiple edges between two
nodes, and the use of grids in which a grid is overlayed on a
planimetric map. Both can result in dense graphs, which require
edge reduction techniques to make their analysis tractable. In
practice such reductions rely on enforcing constraints on feasible
edges in order to build a specific path. Studies making use of such
graphs are provided by \cite{Ayed5}, \cite{Ayed11}, \cite{Ayed13} and
\cite{Fragouli2002}. Hypergraphs are another type of graph used in
some articles. In graph theory, a hypergraph is a generalization
of a graph, where edges can connect any number of vertices,
\cite{D944}. In the multimodal context, such graphs have found
interesting applications in biology and urban transportation.
Sample papers using hypergraphs to represent MMN are yielded by \cite{D933},
\cite{D964} and \cite{D947}.

In effect, the weighted graph approach only utilizes mode
information during the application of constraints, removing the
multimodal traits from the network during modelling.
The analysis in this approach is very application-dependent
as it relies on applying application-specific constraints.

\subsection{The multi--weighted graph approach}\label{SubSe:MultiweightedGraphApproach}

This approach has been extensively utilized for the Multicriteria
Shortest Path Problem (MSPP) which has become an fruitful branch
of research since the 1980s, see \cite{List1} and \cite{Soro2008}
for a complete review. Basically, the approach assigns multiple
weights to each edge. In particular, the bicriteria shortest path
problem assigns two weights to each edge, such as cost and
time.

Optimality in the multi--weighted graph approach is commonly
established by the use of a partial order relation which results
in a Pareto optimal set of paths that are candidates for the
sought shortest path. There is little literature that directly
applies multi--weighted graphs for modelling MMN, but the goal of
MSPP is essentially the same as for the shortest path problem in MMN.
Although articles developing MSPP formulations for MMN can be identified,
they preferentially use partial orders to optimize route choice
decisions mainly based on cost and time, leaving the mode options
as an outcome of the optimal route.
The tractability of the MSPP is inextricably connected with the
cardinality of the Pareto set according to \cite{Muller2006}.
\cite{Hansen_1980} showed in his paper that this cardinality is exponential
in the worst case, although \cite{Loui1983} points out that Pareto
sets for some graphs with multidimensional weights have polynomial
average case cardinalities.
Constraints are applied during analysis to make the problem tractable,
resulting in a Pareto set with manageable cardinality.
MMN models whose mainstay is a multi--weighted graph can be found
in papers by \cite{Andro2009}, \cite{D95} and \cite{D953}.

The paper is organized as follows.
In Section 2 a fourth approach for modeling MMN is introduced and
compared with the three previous approaches.
An algorithm is described in Section 3 that is used in the paper
to compute shortest paths in weighted coloured--edge graphs.
In Section 4 the model is experimentally studied
to demonstrate the tractability of the approach.
Moreover, the algorithm is applied to a real multimodal network
in order to assess its practical applicability.
Finally, Section 5 provides conclusions about this research.

\section{Weighted coloured--edge graph approach}

All the approaches described in Section \ref{Se:Introduction} are heavily
application specific and do not actually utilize the multimodal nature of
a network. In this paper an approach to model and analyze multimodal networks
is introduced. In essence, such approach uses a weighted graph in which edges are
endowed with two attributes: a positive weight and a discrete variable
called colour.

A weighted coloured--edge graph $G=(V,E,\omega,\lambda)$ consists of a
directed graph $(V,E)$ with vertex set $V$ and edge set $E$,
a weight function $\omega\colon E\rightarrow\mathbb{R^+}$ on
edges, and a colour function $\lambda\colon E\rightarrow M$ on edges,
where $M$ is a set of colours.
Typically $M$ is taken as a finite set with $k=|M|$.
Associated to each edge $e_{uv} \in  E$ from vertex $u$ to
vertex $v$ there is a positive weight $\omega(e_{uv})$ and a
colour $\lambda (e_{uv})$. For any colour $i \in M$ and for
any path $p_{uv} = \{e_{x_0x_1},e_{x_1x_2},e_{x_2x_3},\ldots,e_{x_{l-1}x_l}\}$
between two vertices $u=x_0$ and $v=x_1$, where each $x_i \in V$, the path weight $\omega_i
(p_{uv})$ in colour $i$ is defined as
$\omega_i(p_{uv}) = \textstyle \sum_{e_{x_ix_{i+1}} \in p, \; \lambda(e_{x_ix_{i+1}})=i} \omega(e_{x_ix_{i+1}})$.
The total path weight is represented as a $k$-tuple
$(\omega_1(p_{uv}),\ldots,$ $\omega_i(p_{uv}),\ldots,\omega_k(p_{uv}))$,
giving the total weight of the path in each colour.

Note that there is no restriction placed on the number of edges
$e_{uv}$ from a vertex $u$ to a vertex $v$. However, in practice
for the shortest path problem attention can be restricted to
weighted coloured--edge graphs for which there is at most one edge
$e_{uv}$ in each colour from $u$ to $v$.

Let $u$ and $v$ be two given vertices of $G$ and let
$\mathscr{P}_{uv}$ be the set of all paths from $u$ (source) to
$v$ (destination) in $G$. A binary relation between two paths $p_{uv}$
and $p'_{uv}$, is defined by $p_{uv} \le p'_{uv}$ if and only if $\omega_i (p_{uv})
\le \omega_i (p'_{uv})$ for all $i$. The relation $\le$ is clearly
reflexive and transitive and gives a partial order on the
$k$-tuple path weights,
but only a preorder on the paths themselves as multiple paths might have
the same total path weight.

Let $\mathcal{M}_{uv} = \{p_{uv} \in \mathscr{P}_{uv} \mid \forall p'_{uv}
\in \mathscr{P}_{uv}$ with $\omega(p'_{uv}) \ne \omega(p_{uv}), \exists i
\le k$ such that $\omega_i (p_{uv}) < \omega_i (p'_{uv})\}$ be the set of
Pareto optimal paths joining vertices $u$ and $v$. This set has an
important characteristic: for any $p_{uv} \in \mathcal{M}_{uv}$, it is
impossible to determine a path $p^\prime_{uv}$ from $u$ to $v$ which has
smaller weight than $p_{uv}$ in some of its $k$ colours without at least
one of the other weights being larger, analogously to \cite{List16}.

From the above definitions it is apparent that the concept of
a weighted coloured--edge graph with $k$ colours can equivalently be formulated
as a multiweighted multigraph where each edge is assigned a $k$-tuple of non-negative
weights $\left(w_{c_1},\dots,w_{c_i},\dots,w_{c_k}\right)$ and exactly one $w_{c_i}>0$.
However, multiweighted graphs are mostly used in multicriteria optimization applications
where the weight components correspond to quantities to be optimized, such as cost and time,
so edges typically contribute toward more than just one quantity.
For this reason multiweighted graphs whose edge weights are zero in all but one component
have not received attention in the literature.

Shortest path analysis in the weighted coloured-edge graph approach
is seen in this paper to typically be tractable without the need
to apply any application-specific heuristics or constraints,
so can be considered a general tool for the study of multimodal networks.
Application-specific considerations can still be applied to the
resulting set $\mathcal{M}_{uv}$, or a post-optimal analysis undertaken on it.
One facet of this model is that it can be directly applied to
multigraph applications, such as transportation networks where there are multiple
transportation means between two locations, communication networks where there
are multiple links or choice of communication protocols between nodes,
or epidemic models which have multiple paths of infection.

However, focusing attention on only the Pareto optimal paths limits the approach to
shortest path problems where just the summed contribution of each colour
is important, and where any measure of optimality is presumed to be
an increasing (linear or non-linear) function of the summed contribution
in each colour.
For instance, the approach presumes in a transportation network that
the optimal path (such as least cost, time, or distance)
is some application-specific increasing function of the total weight
in each transportation means, or that the user can apply some application-specific
criteria to select a preferred path from the Pareto set once it has been determined.

The approach can be adapted for path constraints
such as restricting the number of hops or the number of mode changes by
slightly enhancing the algorithm used to determine the Pareto set.
For instance, besides using colours to represent the different transportation means,
an additional colour can be used to count the number of edges in a path
as the path is being built during the analysis
and/or to count the number of transfers from one means of transportation to another.

Optimization problems that utilize models similar to weighted coloured--edge graphs
have received little attention in the literature. \cite{Climaco2010} experimentally
studied the number of spanning trees in a weighted graph whose edges are labelled
with a colour. In that work, weight and colour are two criteria to be both minimised
and the proposed algorithm generates a set of non--dominated spanning trees.
The computation of coloured paths in a weighted coloured--edge graph is investigated
in \cite{Xu2009}. The main feature of their approach is a graph reduction technique
based on a priority rule. This rule basically transforms a weighted coloured--edge
multidigraph into a coloured--vertex digraph by applying algebraic operations to the
adjacency matrix. Additionally, the authors provide an algorithm to identify coloured
source--destination paths. Nevertheless, the algorithm is not intended
for general instances because its input is a unit weighted colored multidigraph and only
paths not having consecutive edges equally coloured are considered.

This paper investigates the feasibility of the approach as a general tool for
multimodal networks by determining the cardinality of the Pareto optimal set
$\mathcal{M}_{uv}$ for many diverse networks.
It shows that the cardinality is typically a very low order polynomial function
of the size of the network, and demonstrates that even dense multimodal graphs
with hundreds of thousands of edges can be feasibly analyzed using this approach,
without the need for any reduction techniques.
In fact, it is seen that the number of modes $k$ is more of a limiting
factor of the approach than is the number of vertices or edges in the graph.

\section{Algorithm for determining Pareto Optimal Sets}\label{Se:Algorithm}

To experimentally study the feasibility of using weighted coloured--edge graphs
for multimodal networks an algorithm that determines $\mathcal{M}_{uv}$
is required.
A simple generalization of Dijkstra's algorithm from unimodal networks
has been developed for the purposes of this paper,
although more efficient algorithms might be investigated in the future.
The classic Dijkstra's algorithm for solving the single-source shortest
path problem in unimodal networks uses a priority queue $Q$ to store
shortest path estimates from a fixed source vertex $s$ to each vertex $v$
in the network until the shortest path to $v$ is determined.
Since the weights of any paths $p_{sv}$ from $s$ to $v$ are linearly ordered
there is only at most one shortest path estimate in the queue at a time for each
vertex $v$.
At the start of each iteration of the algorithm the shortest path estimate
at the front of the queue is the actual shortest path to one of the vertices
in the network.

In a weighted coloured--edge graph Dijkstra's algorithm must be slightly
generalised to handle weights of paths being partially ordered rather
than linearly ordered.
A priority queue $Q$ can again be used to store shortest path estimates
with the requirement that if a path $p_{sv}$ from $s$ to $v$ has smaller
weight than another path $p^\prime_{sv}$ then it must appear earlier in the queue.
Although the results presented in this paper use such a simple queue instead
of a more sophisticated non-linear data structure the performance
of the algorithm is seen to be surprisingly good.
As in the classical Dijkstra's algorithm the weighted coloured-edge version of
the algorithm takes as input a network $G$ and a source vertex $s$.
It commences at $s$ with the empty path $p_{ss}$ and relaxes each edge that is incident
from the source vertex $s$, adding the single edge paths to the queue.
At the front of the queue will be a shortest path estimate $p_{sv}$
to some vertex $v$ adjacent to $s$.
Since all weights are positive in the
network $p_{sv}$ must have minimal weight amongst paths from $s$ to $v$
(although it might not be the only minimal path from $s$ to $v$ in the queue),
so $p_{sv}$ is added to the set $\mathcal{M}_{sv}$ and removed from the queue.
The algorithm then relaxes all the edges incident to $v$,
extending the path $p_{sv}$ by each edge to a path $p^\prime_{su}=p_{sv}\cup\{e_{vu}\}$,
adding those extended paths $p^\prime_{su}$ to the queue that have minimal weight
amongst paths from $s$ to $u$, and removing from the queue any path $p^{\prime\prime}_{su}$
from $s$ to $u$ that has greater weight than $p^\prime_{su}$.
The algorithm repeats itself until the queue is empty, producing as output
the Pareto optimal set $\mathcal{M}_{sv}$ for each vertex $v$ in the network.
Figure \ref{pseudo} describes the pseudocode of the algorithm
using the notation developed by \cite{cormen01}.
\begin{figure}[t]
\begin{small}
\begin{codebox}
\Procname{$\proc{Multimodal-Dijkstra}(G, s)$}
\li \Comment Initially no Pareto optimal paths known
\li \For each vertex $v$
\li \Do $\mathcal{M}_{sv} \gets \emptyset$
    \End
\li \Comment Create a queue $Q$ to hold shortest path estimates during processing
\li $Q \gets \emptyset$
\li add the empty path $p_{ss}$ from $s$ to $s$ into $Q$
\li \While $Q \ne \emptyset$
\li \Do remove the path $p_{sv}$ at front of $Q$ that has some end vertex $v$
\li     add the path $p_{sv}$ to $\mathcal{M}_{sv}$
\li     \Comment Relax the edges incident from $v$
\li     \For each edge $e_{vu}$ incident from $v$ to a vertex $u$ not in $p_{sv}$
\li     \Do \Comment Extend the path $p_{sv}$ by the edge $e_{vu}$
\li         $p^\prime_{su} \gets p_{sv}\cup\{e_{vu}\}$
\li         \If $p^\prime_{su}$ has minimal weight in $Q$ from $s$ to $u$
\li         \Then add $p^\prime_{su}$ to $Q$
\li             \Comment Remove any paths no longer minimal in $Q$
\li             \For each $p^{\prime\prime}_{su}\in Q$ with
                   $\omega\left(p^{\prime\prime}_{su}\right)>\omega\left(p^\prime_{su}\right)$
\li             \Do remove the path $p^{\prime\prime}_{su}$ from $Q$
                \End
            \End
        \End
    \End
\li \Return $\mathcal{M}_{sv}$
\end{codebox}
\end{small}
\caption{Pseudocode of the algorithm}
\label{pseudo}
\end{figure}

The number of relaxation steps is an important indicator of the algorithm's
order, so besides finding $\mathcal{M}_{sv}$ the experiment discussed in
Section \ref{Se:ExperimentalStudy} also track the number of paths
$p^\prime_{su}$ processed by the algorithm.

As an example of an application of the weighted
coloured--edge graph approach, the algorithm is run with a
multimodal network from \cite{Lozano2001} starting at source vertex $0$.
Figure \ref{fig00} shows the network which has 21 vertices,
51 edges and 4 different transport choices
(bus, metro, private and transfer).
\begin{figure}
\begin{tiny}
$\scalebox{0.9}{\begin{psmatrix}[colsep=1.2cm,rowsep=0.4cm,mnode=circle,mnodesize=5pt]
&&&\makebox[0.3cm]{6}&&\makebox[0.3cm]{11}\\
&&\makebox[0.3cm]{4}\\
\makebox[0.3cm]{0}&\makebox[0.3cm]{1}&&&&&&\makebox[0.3cm]{18}&&\makebox[0.3cm]{20}\\
&&&&&&\makebox[0.3cm]{16}\\
&&&\makebox[0.3cm]{9}&&\makebox[0.3cm]{13}\\
&&&&&\makebox[0.3cm]{7}&&\makebox[0.3cm]{12}\\
&&&&&&&&&&\makebox[0.3cm]{19}\\
&&&\makebox[0.3cm]{2}\\
&&&&&\makebox[0.3cm]{10}&&\makebox[0.3cm]{14}\\
&&&&&&&\makebox[0.3cm]{8}\\
&&&&&&\makebox[0.3cm]{5}\\
&&&&&\makebox[0.3cm]{3}\\
&{1}&{2}&[mnode=none]{bus}&&&&&&&\makebox[0.3cm]{17}\\
&{1}&{2}&[mnode=none]{metro}&&&&&&\makebox[0.3cm]{15}\\
&{1}&{2}&[mnode=none]{private}\\
&{1}&{2}&[mnode=none]{transfer}
\ncline[linewidth=3.1pt]{->}{3,1}{3,2}\ncput*{15}
\ncline[linewidth=3.1pt]{->}{3,2}{2,3}\ncput*{10}
\ncline[linewidth=3.1pt]{->}{3,2}{5,4}\ncput*{16}
\ncline[linewidth=3.1pt]{->}{2,3}{1,4}\ncput*{12}
\ncline[linewidth=3.1pt]{->}{1,4}{1,6}\ncput*{12}
\ncline[linewidth=3.1pt]{->}{5,4}{5,6}\ncput*{7}
\ncline[linewidth=3.1pt]{->}{5,4}{1,6}\ncput*{8}
\ncline[linewidth=3.1pt]{->}{1,6}{4,7}\ncput*{10}
\ncline[linewidth=3.1pt]{->}{1,6}{3,8}\ncput*{15}
\ncline[linewidth=3.1pt]{->}{5,6}{4,7}\ncput*{13}
\ncline[linewidth=3.1pt]{->}{4,7}{3,8}\ncput*{3}
\ncline[linewidth=3.1pt]{->}{3,8}{3,10}\ncput*{11}
\ncline[linewidth=1.95pt]{->}{8,4}{9,6}\ncput*{22}
\ncline[linewidth=1.95pt]{->}{6,6}{6,8}\ncput*{7}
\ncline[linewidth=1.95pt]{->}{9,6}{9,8}\ncput*{5}
\ncline[linewidth=1.95pt]{->}{6,8}{7,11}\ncput*{19}
\ncline[linewidth=1.95pt]{->}{9,8}{6,8}\ncput*{5}
\ncline[linewidth=1.95pt]{->}{7,11}{3,10}\ncput*{4}
\ncarc[arcangle=-40]{->}{3,1}{12,6}\ncput*{5}
\ncline{->}{12,6}{11,7}\ncput*{7}
\ncline{->}{11,7}{10,8}\ncput*{9}
\ncline{->}{10,8}{14,10}\ncput*{20}
\ncline{->}{14,10}{13,11}\ncput*{2}
\ncarc[arcangle=-30,linestyle=dashed]{<->}{3,2}{12,6}\ncput*{4}
\ncline[linestyle=dashed]{<->}{3,2}{8,4}\ncput*{3}
\ncline[linestyle=dashed]{<->}{8,4}{12,6}\ncput*{4}
\ncarc[arcangle=-29,linestyle=dashed]{<->}{2,3}{11,7}\ncput*{4}
\ncline[linestyle=dashed]{<->}{5,4}{9,6}\ncput*{2}
\ncarc[arcangle=-30,linestyle=dashed]{<->}{1,4}{10,8}\ncput*{4}
\ncline[linestyle=dashed]{<->}{1,4}{6,6}\ncput*{2}
\ncline[linestyle=dashed]{<->}{6,6}{10,8}\ncput*{4}
\ncline[linestyle=dashed]{<->}{5,6}{9,8}\ncput*{2}
\ncarc[arcangle=30,linestyle=dashed]{<->}{5,6}{14,10}\ncput*{4}
\ncline[linestyle=dashed]{<->}{9,8}{14,10}\ncput*{3}
\ncarc[arcangle=-33,linestyle=dashed]{<->}{1,6}{6,8}\ncput*{3}
\ncarc[arcangle=30,linestyle=dashed]{<->}{4,7}{13,11}\ncput*{2}
\ncline[linestyle=dashed]{<->}{3,8}{7,11}\ncput*{1}
\ncline[linewidth=3.1pt]{->}{13,2}{13,3} 
\ncline[linewidth=1.95pt]{->}{14,2}{14,3}
\ncline{->}{15,2}{15,3} 
\ncline[linestyle=dashed]{->}{16,2}{16,3} 
\end{psmatrix}}$
\end{tiny}
\caption{Multimodal network from \cite{Lozano2001}}
\label{fig00}
\end{figure}
The algorithm commences with just the empty path $p_{00}$ on the queue and
relaxes two edges: $e_{01}$ with weight
$(\mbox{bus}, \mbox{metro}, \mbox{private}, \mbox{transfer})= (15,0,0,0)$,
and $e_{03}$ with weight $(0,0,5,0)$, which are both added to the queue.
Since the two weights are incomparable, either could be at the front of the queue,
so the next iteration of the algorithm either adds the path $p_{01}=\{e_{01}\}$
to $\mathcal{M}_{01}$ and relaxes the four edges incident to vertex $1$ by extending the
path $p_{01}$ by each, or else adds the path $p_{03}=\{e_{03}\}$ to $\mathcal{M}_{03}$
and relaxes the three edges incident to vertex $3$ by extending the path $p_{03}$ by each.
Continuing in this way the Pareto optimal set $\mathcal{M}_{0v}$ is obtained for
each vertex $v$ in the network, resulting in $52$ Pareto optimal paths from vertex $0$
to vertex $20$ whose weights are listed in Table \ref{tab:a}.
\begin{table}
\caption{Pareto set for network with 21 vertices and 51
edges.}\label{tab:a} \vspace{-0.1 cm}
\begin{center}\begin{scriptsize}
\begin{tabular}{|c|cccc|c|}\hline
    & \multicolumn{4}{c|}{Transport Choice Cost}& Cost as per\\ \cline{2-5}
Path Number & Bus & Metro & Private & Transfer & {\tiny{\cite{Lozano2001}}}\\ \hline
  1 & 25 & 4  & 21 & 5  & 55 \\
  2 & 0  & 30 & 21 & 4  & 55 \\
  3 & 32 & 9  & 5  & 9  & 55 \\
  4 & 13 & 11 & 21 & 8  & 53 \\
  5 & 24 & 0  & 36 & 10 & 70 \\
  6 & 11 & 26 & 21 & 5  & 63 \\
  7 & 21 & 26 & 5  & 7  & 59 \\
  8 & 50 & 19 & 0  & 4  & 73 \\
  9 & 41 & 4  & 2  & 7  & 54 \\
 10 & 8  & 45 & 5  & 9  & 67 \\
 11 & 3  & 4  & 43 & 3  & 53 \\
 12 & 13 & 4  & 36 & 11 & 64 \\
 13 & 10 & 30 & 14 & 12 & 66 \\
 14 & 25 & 26 & 2  & 12 & 65 \\
 15 & 26 & 0  & 34 & 10 & 70 \\
 16 & 3  & 31 & 7  & 10 & 51 \\
 17 & 16 & 4  & 41 & 5  & 66 \\
 18 & 47 & 9  & 0  & 5  & 61 \\
 19 & 31 & 31 & 0  & 6  & 68 \\
 20 & 14 & 0  & 43 & 2  & 59 \\
 21 & 23 & 45 & 0  & 8  & 76 \\
 22 & 52 & 4  & 0  & 1  & 57 \\
 23 & 24 & 7  & 21 & 7  & 59 \\
 24 & 25 & 11 & 12 & 10 & 58 \\
 25 & 19 & 9  & 7  & 12 & 47 \\
 26 & 32 & 22 & 5  & 6  & 65 \\
 27 & 16 & 31 & 5  & 7  & 59 \\
 28 & 29 & 27 & 2  & 8  & 66 \\
 29 & 36 & 0  & 21 & 4  & 61 \\
 30 & 15 & 4  & 34 & 11 & 64 \\
 31 & 14 & 27 & 7  & 9  & 57 \\
 32 & 12 & 23 & 21 & 7  & 63 \\
 33 & 63 & 0  & 0  & 0  & 63 \\
 34 & 39 & 23 & 0  & 3  & 65 \\
 35 & 24 & 23 & 5  & 7  & 59 \\
 36 & 36 & 26 & 0  & 6  & 68 \\
 37 & 12 & 30 & 12 & 6  & 60 \\
 38 & 10 & 26 & 7  & 13 & 56 \\
 39 & 18 & 31 & 2  & 9  & 60 \\
 40 & 27 & 0  & 41 & 4  & 72 \\
 41 & 26 & 4  & 7  & 11 & 48 \\
 42 & 29 & 0  & 38 & 6  & 73 \\
 43 & 52 & 0  & 2  & 6  & 60 \\
 44 & 22 & 30 & 5  & 6  & 63 \\
 45 & 34 & 9  & 2  & 8  & 53 \\
 46 & 48 & 0  & 5  & 4  & 57 \\
 47 & 37 & 4  & 5  & 5  & 51 \\
 48 & 37 & 30 & 0  & 2  & 69 \\
 49 & 36 & 7  & 12 & 9  & 64 \\
 50 & 18 & 4  & 38 & 7  & 67 \\
 51 & 27 & 27 & 5  & 6  & 65 \\
 52 & 37 & 0  & 7  & 10 & 54 \\
 \hline
\end{tabular}
\end{scriptsize}\end{center}
\end{table}
Depending on the application, constraints or heuristics can then be applied to
the $52$ paths to select a path preferred by the user.
Using just a simple priority queue data structure the generalised
Dijkstra's algorithm can determine $\mathcal{M}_{0v}$ for all $21$ vertices $v$
within approximately $10\mbox{ms}$.
The article of \cite{Lozano2001} instead uses a weighted graph approach with
application-specific constraints and a simple cost function which adds the
weights in each mode together to get a single valued total weight,
resulting in the paths numbered $2$, $25$, $33$, $47$ in the table.

\begin{figure}
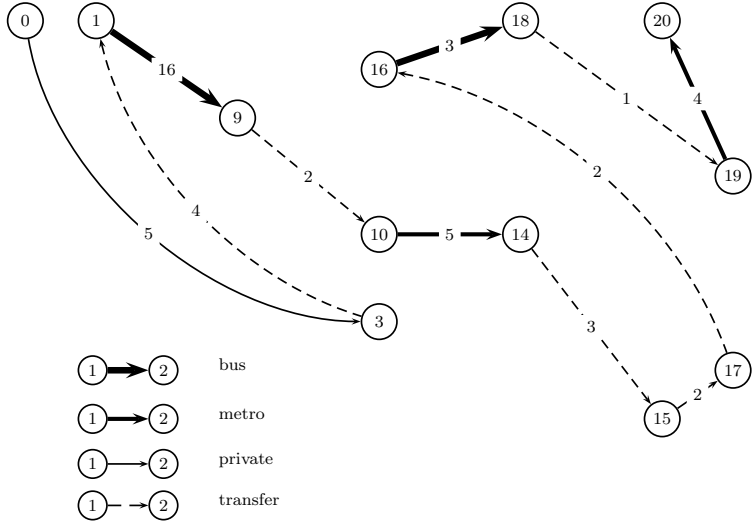

\begin{center}
\begin{scriptsize}
$\scalebox{0.8}{\begin{psmatrix}[colsep=1cm,rowsep=0.15cm,mnode=circle,mnodesize=5pt]
&&&&&[mnode=none]\\
&&[mnode=none]\\
\makebox[0.3cm]{0}&\makebox[0.3cm]{1}&&&&&&\makebox[0.3cm]{18}&&\makebox[0.3cm]{20}\\
&&&&&\makebox[0.3cm]{16}\\
&&&\makebox[0.3cm]{9}&&[mnode=none]\\
&&&&&&&[mnode=none]\\
&&&&&&&&&&\makebox[0.3cm]{19}\\
&&&[mnode=none]\\
&&&&&\makebox[0.3cm]{10}&&\makebox[0.3cm]{14}\\
&&&&&&&[mnode=none]\\
&&&&&&[mnode=none]\\
&&&&&\makebox[0.3cm]{3}\\
&{1}&{2}&[mnode=none]{bus}&&&&&&&\makebox[0.3cm]{17}\\
&{1}&{2}&[mnode=none]{metro}&&&&&&\makebox[0.3cm]{15}\\
&{1}&{2}&[mnode=none]{private}\\
&{1}&{2}&[mnode=none]{transfer}
\ncline[linewidth=3.1pt]{->}{3,2}{5,4}\ncput*{16}
\ncline[linewidth=3.1pt]{->}{4,6}{3,8}\ncput*{3}
\ncline[linewidth=1.95pt]{->}{9,6}{9,8}\ncput*{5}
\ncline[linewidth=1.95pt]{->}{7,11}{3,10}\ncput*{4}
\ncarc[arcangle=-40]{->}{3,1}{12,6}\ncput*{5}
\ncline{->}{14,10}{13,11}\ncput*{2}
\ncarc[arcangle=-30,linestyle=dashed]{<-}{3,2}{12,6}\ncput*{4}
\ncline[linestyle=dashed]{->}{5,4}{9,6}\ncput*{2}
\ncline[linestyle=dashed]{->}{9,8}{14,10}\ncput*{3}
\ncarc[arcangle=30,linestyle=dashed]{<-}{4,6}{13,11}\ncput*{2}
\ncline[linestyle=dashed]{->}{3,8}{7,11}\ncput*{1}
\ncline[linewidth=3.1pt]{->}{13,2}{13,3} 
\ncline[linewidth=1.95pt]{->}{14,2}{14,3}
\ncline{->}{15,2}{15,3} 
\ncline[linestyle=dashed]{->}{16,2}{16,3} 
\end{psmatrix}}$
\end{scriptsize}
\vspace{0.5 cm}
\caption{Path $25$ used for a post--optimal analysis}
\label{path13}
\end{center}
\end{figure}

Note that a Pareto set permits a post--optimal analysis to be carried out
provided that the total cost is presumed to be an increasing
function of the weight in each mode.
For instance, suppose in the example network that the edge weights
represent distance and for simplicity presume the cost is one dollar
per unit distance for each means of transport.
A natural optimization question could be how much the
unit cost associated to a particular mode could be increased or
decreased with the current optimal path remaining optimal.
As an illustration, path $25$ which has the edges
$\left\{e_{03}, e_{31}, e_{1\,9}, e_{9\,10}, e_{10\,14}, e_{14\,15},
e_{15\,17}, e_{17\,16}, e_{16\,18}, e_{18\,19}, e_{19\,20}\right\}$
and shown in Figure \ref{path13} has least total cost $\$47$,
but from the Pareto set it is easily seen that an
increase of over $20\%$ in the relative metro costing would make
path $41$ with edges $\left\{e_{03},e_{31},e_{19},e_{9\,13},e_{13\,15},e_{15\,17},
e_{17\,16},e_{16\,18},e_{18\,19},e_{19\,20}\right\}$ a better choice, or
a $25\%$ increase in bus prices would make
path $16$ with edges $\left\{e_{03},e_{32},e_{2\,10},e_{10\,14},e_{14\,15},
e_{15\,17},e_{17\,16},e_{16\,18},e_{18\,19},e_{19\,20}\right\}$ better.

This example demonstrates that the Multimodal Dijkstra's algorithm
can quickly calculate the Pareto set without the need to assign relative
costs for the different modes. Then alternative cost functions can
be evaluated on just the paths in the Pareto set or a post--optimal
analysis conducted without ever having to rerun the algorithm on the network.

\section{Experimental Study}\label{Se:ExperimentalStudy}

In this section the weighted coloured--edge graph approach is applied to
multimodal networks in two different scenarios.
Firstly, the cardinality of $\mathcal{M}_{uv}$ is analyzed for
complete graphs. Secondly, the approach is applied to a real multimodal
network. This network corresponds to the transportation system of France
and considers four transport choices.

\subsection{Number of processing paths and cardinality of $\mathcal{M}_{uv}$}

The objective here is to identify general patterns for the number
of processing paths and $\mathcal{M}_{uv}$ cardinality.
In this test a weighted complete multigraph is taken as
input so that each analytical scenario is generated by fixing
values for $n=|V|$ and $k=|M|$. Such a graph is characterized by
having $kn(n-1)$ edges and the maximum number of possible paths
$|\mathscr{P}_{uv}| = \tiny{\sum_{j=0}^{n-2}\bin{n-2}{j}k^{j+1}j!}$
for $v\neq u$, which has factorial order $O\left(k^{n-1}(n-2)!\right)$.
Specifically, the algorithm is run for complete multigraphs with
$k=2,3,4,5$ colours and $n$ between 20 and 200 vertices.
Random edge weights are generated by means of a continuous uniform
distribution of positive weights.

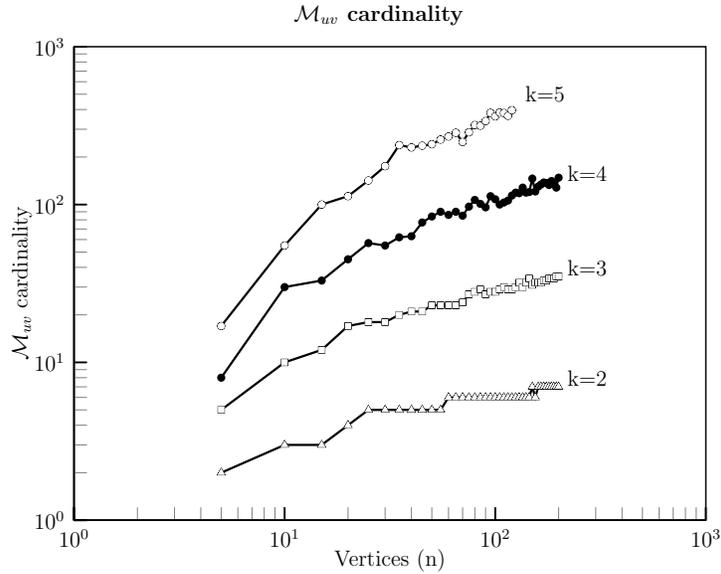
\begin{figure}[t]
\psset{xunit= 4 cm,yunit= 3 cm}
\def\datapsII{
5   2
10  3
15  3
20  4
25  5
30  5
35  5
40  5
45  5
50  5
55  5
60  6
65  6
70  6
75  6
80  6
85  6
90  6
95  6
100 6
105 6
110 6
115 6
120 6
125 6
130 6
135 6
140 6
145 6
150 7
155 6
160 7
165 7
170 7
175 7
180 7
185 7
190 7
195 7
200 7
}
\def\datapsIII{
5   5
10  10
15  12
20  17
25  18
30  18
35  20
40  21
45  21
50  23
55  23
60  23
65  23
70  24
75  27
80  28
85  29
90  27
95  28
100 28
105 29
110 30
115 29
120 29
125 30
130 32
135 30
140 32
145 34
150 31
155 32
160 32
165 32
170 33
175 33
180 34
185 34
190 34
195 35
200 35
}
\def\datapsIV{
5   8
10  30
15  33
20  45
25  57
30  55
35  62
40  63
45  77
50  84
55  90
60  86
65  90
70  85
75  97
80  107
85  101
90  96
95  113
100 108
105 100
110 103
115 106
120 114
125 119
130 118
135 128
140 119
145 120
150 146
155 121
160 130
165 134
170 138
175 137
180 133
185 141
190 135
195 128
200 148
}
\def\datapsV{
5   17
10  55
15  100
20  113
25  142
30  175
35  238
40  230
45  236
50  241
55  258
60  270
65  285
70  249
75  287
80  319
85  316
90  338
95  381
100 362
105 383
110 380
115 365
120 396
}
\scalebox{0.7}{
\begin{pspicture}(17,3.4)(-1,0)
  \pstScalePoints(1,1){log}{log}
  \psaxes[ axesstyle=frame, xylogBase=10, logLines=all, tickwidth=1pt,
    ticksize=0 7pt, subtickwidth=0.5pt, subticks=10](3,3)
  \listplot[showpoints=true, linecolor=black, linewidth=1.2pt,
    dotsize=2pt 2, dotstyle=triangle ]{\datapsII}
  \listplot[showpoints=true, linecolor=black,
    linewidth=1.2pt, dotsize=2pt 2, dotstyle=square ]{\datapsIII}
  \listplot[showpoints=true, linecolor=black, dotstyle=*,
    linewidth=1.2pt, dotsize=2pt 2]{\datapsIV}
  \listplot[showpoints=true, linecolor=black, dotstyle=o,
    linewidth=1.2pt, dotsize=2pt 2]{\datapsV}
  \uput{1.0cm}[270](1.5,0.15){Vertices (n)}
  \uput{1.2cm}[180]{90}(0.1,1.5){$\mathcal{M}_{uv}$ cardinality}
  \uput[0](1,3.2){\bf{$\mathcal{M}_{uv}$ cardinality}}
  \uput[0](2.1,2.7){k=5}
  \uput[0](2.3,2.2){k=4}
  \uput[0](2.3,1.6){k=3}
  \uput[0](2.3,0.9){k=2}
\end{pspicture} }
\vspace{0.2cm}
\caption{Cardinality of $\mathcal{M}_{uv}$ for random weighted coloured--edge graphs
with different number of colours $k$.} \label{exp01}
\end{figure}

Figure \ref{exp01} depicts the patterns followed by
$\mathcal{M}_{uv}$ cardinality. The figure uses a logarithmic scale for vertical as well as horizontal
axes to demonstrate the average case polynomial behavior as $n$ increases.
Table \ref{orders1} provides the numerical orders determined for different $k$ values.
These results demonstrate not only that the Pareto optimal set is calculated in
polynomial time but that the resulting set requiring further analysis
grows very slowly as a function of $n$.
The results resemble ideas presented in
\cite{Bentley1978} and \cite{Muller2001}, suggesting the
applicability of the model in real multimodal network scenarios,
even when the networks are dense and without having to apply
network reduction techniques or heuristics.
\begin{table}
\caption{Order of processing paths and $\mathcal{M}_{uv}$
\@cardinality for several $k$ values}\label{orders1}
\vspace{-0.08in}
\begin{center}
\begin{tabular}{lcc}\hline
  $k$ & \hspace{1.0cm}Processing paths ($p^\prime_{su}$)&\hspace{1.0cm}$\mathcal{M}_{uv}$ cardinality \\ \hline
  2   & $O(n^{1.28})$ & $O(n^{0.19})$\\
  3   & $O(n^{1.37})$ & $O(n^{0.32})$\\
  4   & $O(n^{1.52})$ & $O(n^{0.46})$\\
  5   & $O(n^{1.64})$ & $O(n^{0.61})$\\ \hline
\end{tabular}
\end{center}
\end{table}


\subsection{Performance in a real multimodal network}

The approach is now tested on a large multimodal network. In this test,
largeness is in the sense of number of vertices and edges. The selected network scenario
corresponds to the multimodal transportation system of France being one of the
largest networks in Europe. The multimodal network was obtained from vector data
information retrieved from a public GIS library, \cite{cloud2010}.

The network dataset for each transport choice was firstly processed
in ArcGIS to make it suitable for computation. ArcGIS is a Windows
platform application for the analysis and processing of
vector geographic information system data. This application has a network analysis
extension that permits the identification of junctions and polylines
(see \cite{Burke2002} for a definition) in each transport system. In
addition, ArcGIS also has a macro for the computation
of the adjacency matrix for each system of junctions.

\begin{figure}
\scalebox{0.5}{
\includegraphics[60,0][450,450]{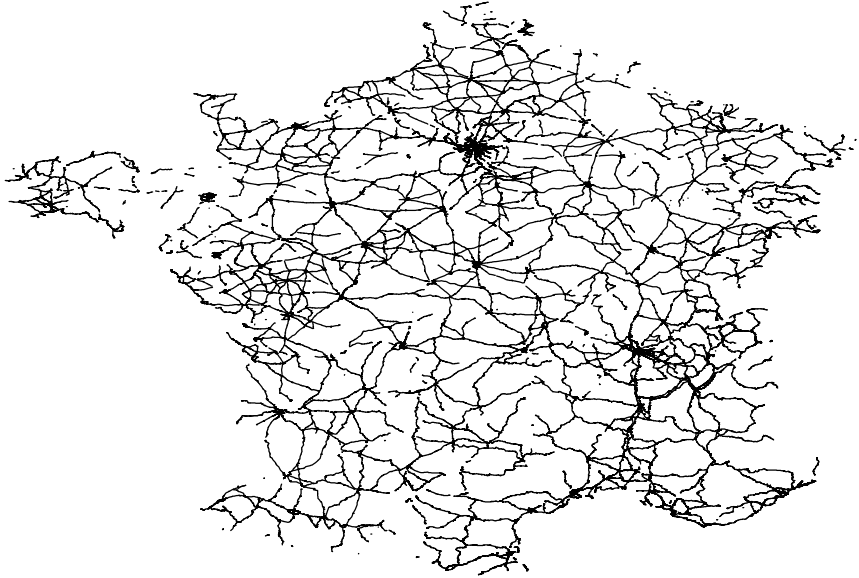}}
\caption{France roadway system.}
\label{FranceRW}
\end{figure}

Table \ref{nets2} summarizes the number of junctions and  edges for each transport
mode as well as some statistics of the networks.
\begin{table}[b]
\caption{Characteristics of France Multimodal Network}\label{nets2}
\vspace{-0.5cm}
\begin{center}
\begin{tabular}{lccccc}\hline
  Modes          & Number of          & Number of & \multicolumn{3}{c}{Edge length}       \\ \cline{4-6}
                 & Junctions          & polylines     & Maximum    & Mean        & Stnd. Dev.  \\ \hline 
  Roadways       & 53562              & 47660     & 0.868028   & 0.010674    & 0.032452     \\
  Railways       & 18671              & 20083     & 1.280264   & 0.014966    & 0.046192     \\
  Motorway       & 7720               & 7432      & 1.221951   & 0.033488    & 0.078485      \\
  Waterways      & 17113	            & 11635     & 3.238686   & 0.032070    & 0.095573       \\
  \hline
\end{tabular}
\end{center}
\label{junctions}
\end{table}
All edge lengths are given in decimal geographic degrees.
Four transport modes comprise the France transport system: road, rail, waterways and
motorways. The road system  mainly consists of primary roads. The rail system is
comprised of common train lines disregarding subway and tram. Waterways are the
channels and rivers used as transportation links. Finally, the motorway system of
France includes toll roads and is considered a different mode of transport in its
own right. As an illustration,
Figure \ref{FranceRW} depicts the France road system.

The construction of the multimodal network requires assembling the data
for the four network modes together. This task is accomplished by an
ad--hoc algorithm coded in the Java language. The code basically takes
two inputs. These are the adjacency matrix of each transport mode network
and a list of minimum interjunction distances in each mode. The latter
is built in ArcGIS by taking a mode junction dataset and applying the
``join and relates'' tool with respect to each other mode junction dataset.
This information facilitates the performance of a subsequent
clustering procedure used inside of the ad--hoc algorithm.

Two parameters need to be specified once the algorithm code is executed. 
A minimum clustering distance (this generates the network vertices)
and a source vertex (a junction number). After entering this information,
the Java code invokes the multimodal Dijkstra's algorithm, reporting at
the end a list with the total number of optimal paths to each vertex
together with two additional variables: the maximum number of paths found
in a particular vertex and the average number of paths.

This dataset was tested by assigning a cluster distances between $0.150$ and $0.115$
decimal degrees ($14$ to $11$ km). The resulting networks together with running times
(minutes) and average number of optimal paths are shown in Table \ref{FranceResults}.
\begin{table}
\caption{Results of France Multimodal Network}\label{FranceResults}
\vspace{-0.1cm}
\begin{center}
\begin{tabular}{cccccc}\hline
  Cluster  &  Number of & Number of & Running & Avg.    & Max                \\
           &  Vertices  & Edges     & Time    & Paths   & Paths              \\ \hline 
  0.150    &  1501      & 4216      & 0.0182   & 33.0930  & 702              \\
  0.140    &  1869      & 5218      & 0.6953   & 208.270  & 3320               \\
  0.130    &  2343      & 6280      & 4.4010   & 404.380  & 14972                \\
  0.120    &  2948      & 7696      & 245.40   & 2013.82  & 37128                  \\
  0.115    &  3108      & 8018      & 823.10   & 3575.03  & 46984                 \\
   \hline
\end{tabular}
\end{center}
\label{FranceResults}
\end{table}
The computations were performed on a standard desktop computer with dual core
2.93 GHz CPU and 8 GB of RAM that was set with the queue version of the
multimodal Dijkstra algorithm.

Although the networks shown in Table \ref{FranceResults} requiring longer
runs of the multimodal Dijkstra's algorithm when the clustering distance
is reduced, it cannot be disregarded that no constraints or reductions were
required for obtaining the results in Table \ref{FranceResults} and the
cardinality of the resulting Pareto sets are quite manageable for any
further analysis.  

\section{Conclusions}

In modelling multimodal networks, current approaches for determining shortest
paths rely on applying application-specific constraints or heuristics to
obtain tractable problems.
This paper introduces a modelling approach that keeps the multimodal
traits of a network by assigning discrete colour attributes to the
edges, uses a partial order to obtain a Pareto set of paths of potential interest,
and avoids the need for reduction techniques.
Although a straightforward approach to modelling networks in which there
are multiple transportation modes, it does appear to give a new perspective
and truly general approach for multimodal networks.
Another feature of this approach is that it results in a Pareto
set, which can be further investigated without rerunning the
algorithm. This opens the door to post--optimal analysis in MMN.

The experimental study analyzing the Pareto set indicates that its
cardinality is typically a low order polynomial for random
uniformly distributed weighted coloured--edge graphs.
Furthermore, the approach can deal with networks as large as the
multimodal transportation system of France. 

\section*{Acknowledgements}
This research is partly supported by Universidad Catolica del Maule,
Talca--Chile, through the project MECESUP--UCM0205.

\bibliographystyle{elsarticle-harv}
\bibliography{biblio1}{}

\end{document}